\documentclass{article}
\pdfoutput=1

\usepackage{arxiv}

\usepackage[utf8]{inputenc} 
\usepackage[T1]{fontenc}    
\usepackage{hyperref}       
\usepackage{url}            
\usepackage{booktabs}       
\usepackage{amsfonts}       
\usepackage{amssymb, amsmath}
\usepackage{nicefrac}       
\usepackage{microtype}      
\usepackage{lipsum}
\usepackage{graphicx}
\usepackage{setspace}
\usepackage{array}
\usepackage{rotating}
\usepackage[authoryear]{natbib}
\graphicspath{{Figures/}}

\newcommand{\mic}{$ft > k\ \times \textrm{ MIC}$}
\newcolumntype{L}[1]{>{\raggedright\let\newline\\\arraybackslash\hspace{0pt}}p{#1}}

\title{Strategies to mitigate bias from time recording errors in pharmacokinetic studies}

\author{
 Hannah L Weeks \\
  Department of Biostatistics\\
  Vanderbilt University Medical Center\\
  Nashville, TN 37203 \\
  \texttt{hannah.l.weeks@vanderbilt.edu} \\
  \And
 Matthew S Shotwell \\
  Department of Biostatistics\\
  Vanderbilt University Medical Center\\
  Nashville, TN 37203 \\
  \texttt{matthew.shotwell@vumc.org} \\
}

\date{October 23, 2023}

\begin{document}

\maketitle

\begin{abstract}
Opportunistic pharmacokinetic (PK) studies have sparse and imbalanced clinical measurement data, and the impact of sample time errors is an important concern when seeking accurate estimates of treatment response. We evaluated an approximate Bayesian model for individualized pharmacokinetics in the presence of time recording errors (TREs), considering both a short and long infusion dosing pattern. We found that the long infusion schedule generally had lower bias in estimates of the pharmacodynamic (PD) endpoint relative to the short infusion schedule. We investigated three different design strategies for their ability to mitigate the impact of TREs: (i) shifting blood draws taken during an active infusion to the post-infusion period, (ii) identifying the best next sample time by minimizing bias in the presence of TREs, and (iii) collecting additional information on a subset of patients based on estimate uncertainty or quadrature-estimated variance in the presence of TREs. Generally, the proposed strategies led to a decrease in bias of the PD estimate for the short infusion schedule, but had a negligible impact for the long infusion schedule. Dosing regimens with periods of high non-linearity may benefit from design modifications, while more stable concentration-time profiles are generally more robust to TREs with no design modifications.
\end{abstract}


\section{Introduction}

Pharmacokinetic studies rely on two key pieces of information: (i) drug concentration measurements and (ii) the time at which those measurements were taken. Individual-level data provides feedback about how a patient is responding to treatment up to the time of measurement, allowing healthcare providers to adaptively modify and individualize treatment. The time at which a blood sample was taken is a critical aspect of this patient data. In order to appropriately model patient-specific concentration-time curves, measurements must be linked to the appropriate time at which blood was drawn (relative to when treatment began). In practice, it is possible for errors to occur in both the blood draw and dose administration times (i.e., times of measurement and treatment), particularly when dealing with opportunistic samples.

Opportunistic sampling designs in observational PK studies collect data through routine care on patients. Unlike a more structured PK study, such as a clinical trial, this type of data is typically unbalanced (i.e., patients have differing numbers of blood draws) and more stochastic (i.e., blood draws can occur at random times throughout treatment). It is also common to have very sparse data for each patient, even fewer than 5 observations per patient \citep{charles2014population}. Thus, the ability to provide estimates that are robust to recording errors in sample times is an important consideration for a PK model to fit opportunistic data.

\subsection{Time recording errors in pharmacokinetic analyses}

The impact of TREs has been looked at previously in population PK studies. \citet{sun1996} found that sampling time errors of more than 20\% the true sample time led to bias in parameter estimates, particularly when only a single dose was administered via intravenous bolus injection. The highest magnitudes of bias were observed in estimates of intra- and inter-subject variability \citep{sun1996}.

\citet{wang1996note} demonstrated that the theoretical impact of Berkson sample time error on bias in the PK parameters was substantial. In contrast to classical ME the Berkson error model indicates that the unobserved (true) sample times are more variable than the observed times \citep{berkson1950there}. In structured PK studies, this assumption typically holds since sample times are fixed according to a scheduled protocol. However, simulation results indicated that under a dense sampling design, this bias was less prevalent than was theoretically expected \citep{wang1996note}. \citet{choi2013} extended these results via a simulation with less dense sampling times for a population PK study. While Berkson error does not necessarily always cause bias in linear regression estimates, they found that it did lead to bias in estimates of the PK parameters. Further, the amount of bias was highest for areas of the concentration-time curve with high curvature (i.e., nonlinearity) \citep{choi2013}.

Comparison across patients can be used to look for outlier concentration values to determine measurements that are likely the result of underlying error \citep{irby2021approaches}. In general, though, it can sometimes be difficult to determine whether an error in the concentration-time curve is due to error in the concentration measurement, error in the sampling time, or both. \citet{irby2021approaches} recommend removing entire observations that are known to contain significant ME in the sample time or measured concentration. However, this is not strictly necessary as statistical approaches such as transform-both-sides models have been shown to reduce the bias in both fixed and random effects estimates of population PK models in the presence of sample time error \citep{choi2013}.

Existing approaches have focused on how to deal with sample time errors in analyses for existing data sets. \citet{stefanski2000measurement} noted the importance of correctly specifying the ME model, stating that incorrect assumptions can potentially be just as problematic as using naive estimates which ignore ME entirely. We turn now to the question of whether the sampling design itself can be modified to provide robustness in the presence of such recording errors. We first characterize the impact of errors in blood draw times and infusion times on model estimates of a pharmacodynamic (PD) endpoint, using the individual-level approximate Bayesian PK model for repeated intravenous infusions of a drug presented in Chapter 3. As TREs are a known risk with PK analyses, particularly when using opportunistic samples, it is important to determine both (i) how the model performs in the presence of errors and (ii) strategies for mitigating the impact of this type of error.

This chapter addresses two main objectives, each of which have their own methods and results subsections. Objective 1 aims to characterize the impact of TREs on estimates of the PD endpoint through simulations. Objective 2 focuses on various proposed policy design changes that attempt to mitigate the impact of TREs on bias in the PD endpoint.

\section{Context and simulation setup}

\subsection{Notation}

First, we introduce some notation that will be used throughout this section. Let $i$ refer to a patient identifier and $j$ refer to a time identifier. Times of blood draws are noted as $t_{ij}$ for the $i^{th}$ patient at the $j^{th}$ time. The notation $c_{ij}$ will refer to the observed concentration for patient $i$ at the corresponding time $t_{ij}$.

We distinguish between true times versus recorded times in the following way: true times $t_{ij}$ correctly reflect the time at which a concentration measurement was taken, while recorded times $t_{ij}^*$ are observed yet error-prone, and deviate from true times by a random amount $\delta$. In other words, $t_{ij}^* = t_{ij} + \delta$, where $\delta \sim \textrm{N}(0, \sigma_{RE})$. All time values are measured in ``hours since administration of the first dose'' (h). In this series of simulation studies, we considered $\sigma_{RE} \in \{0.25, 0.5, 1.0, 1.5\}$, corresponding to a standard deviation (SD) of 15, 30, 60, and 90 minutes, respectively, for the errors $\delta$.

Unless otherwise specified, in the simulation studies and results we compared two estimates: (i) $h(\hat{\boldsymbol{\theta}}_i^*)$ versus (ii) $h(\hat{\boldsymbol{\theta}}_i)$. The values $\hat{\boldsymbol{\theta}}_i^*$ and $\hat{\boldsymbol{\theta}}_i$ reflect the posterior mode estimate of the patient-specific PK parameter based on the recorded (i.e., incorrect) and true times, respectively. That is, given $\hat{\boldsymbol{\theta}}_i$, the posterior for patient $i$, the modes are defined as

\begin{align*}
\hat{\boldsymbol{\theta}}_i^* &= argmax_{\boldsymbol{\theta}}\ \tilde{\pi}_i(\boldsymbol{\theta}_i, \sigma_i | \mathbf{c}_i, \mathbf{t}_i^*) \\
\hat{\boldsymbol{\theta}}_i &= argmax_{\boldsymbol{\theta}}\ \tilde{\pi}_i(\boldsymbol{\theta}_i, \sigma_i| \mathbf{c}_i, \mathbf{t}_i)
\end{align*}

\subsection{Model overview and clinical context}

A brief overview of the model is provided here (see Chapter 3 for details). Using the two compartment model solution $\eta(t_{ij}, \boldsymbol{\theta}_i)$, the individual-level likelihood for patient $i$ is defined as:

\begin{align}
L_i(\boldsymbol{\theta}_i, \sigma_i|\mathbf{c}_i, \mathbf{t}_i) = \prod_{j=1}^{m_i} \phi\bigg(\cfrac{c_{ij} - \eta_i(t_{ij}, \boldsymbol{\theta}_i)}{\sigma_i}\bigg).
\end{align}

The prior $\pi_0(\boldsymbol{\theta}_i)$ is generated such that $\log \boldsymbol{\theta}_i \sim N_4(\boldsymbol{\mu}_0, \boldsymbol{\Sigma}_0)$ and $\log \sigma_i \sim N_1(m_0, s_0)$, where the hyperparameters $\boldsymbol{\mu}_0$, $\boldsymbol{\Sigma}_0$, $m_0$, and $s_0$ are derived from a prior analysis in a similar population of critically ill patients being treated with the antibiotic piperacillin. Using a Laplace approximation, the patient-specific posterior is approximated as

\begin{align}
\tilde{\pi}_i(\log \boldsymbol{\theta}_i, \log \sigma_i) \approx \textrm{N}([\log \hat{\boldsymbol{\theta}}_i,\log \hat{\sigma}_i], [-H_{\hat{\boldsymbol{\theta}}_i}]^{-1})
\end{align}

\noindent using the posterior mode $[\log \hat{\boldsymbol{\theta}}_i,\log \hat{\sigma}_i]$ and posterior Hessian evaluated at the mode $H_{\hat{\boldsymbol{\theta}}_i}$.

The pharmacodynamic endpoint of interest is $h(\boldsymbol{\theta}_i)$ = \mic. This statistic is the fraction of the patient's dosing history spent above a multiple of the minimum inhibitory concentration (MIC), or the minimum concentration of a drug required to suppress infection. Using a multiple of the MIC accounts for this expected attenuation at the infection site (e.g., within an organ or muscle) relative to the site where observed concentration measurements are drawn (the bloodstream). For the drug of interest piperacillin, we set $k=4$ for a $k \times$ MIC threshold of 64 $\mu$g/mL \citep{shotwell2016}.

\section{Objective 1: Assessing sensitivity to time recording errors}

\subsection{Objective 1 methods}

We considered two different infusion schedules of 3 grams of piperacillin administered every 8 hours with five total infusions, for both a short (30 minute) and long (4 hour) infusion schedule. The start of the first infusion was considered 0 hours and all other dose and sample times were defined relative to this start point. Sample vectors of $\boldsymbol{\theta}_i$ were drawn from the prior distribution to represent simulated patients (i.e., treated as each patient's true PK parameters). Two different approaches were taken to characterize the impact of TREs.

For the first approach, we sampled 1000 simulated patients. For each patient and each infusion schedule, we simulated blood draw times $t_{ij} \sim \textrm{Unif}(s_j, e_j)$ for $j = 1,...,5$, where $s_j$ and $e_j$ represent the start and end times of each dose window. In accordance with the infusion schedules, each dose window was 8 hours in length and extended until just before the next dose was administered. For each $t_{ij}$, using the patient's assumed true PK parameter $\boldsymbol{\theta}_i$ we simulated an observed concentration measurement $c_{ij}$ according to the two compartment model solution $\eta(t_{ij}, \boldsymbol{\theta}_i)$ with additive random error. For each patient and infusion schedule, we obtained the following estimates:

\begin{enumerate}
  \item For each time index $j$, the model estimate $h(\hat{\boldsymbol{\theta}}_i)$ based on correct information up to and including time $j$: $\{\mathbf{c}_i,\mathbf{t}_i\}$ where $\mathbf{c}_i = \{c_{i1},...,c_{ij}\}$ and $\mathbf{t}_i = \{t_{i1},...,t_{ij}\}$
  \item For each time index $j$, the model estimates $h(\hat{\boldsymbol{\theta}}_i^*)$ based on error-prone information up to and including time $j$: $\{\mathbf{c}_i,\mathbf{t}_i^*\}$ where $\mathbf{c}_i = \{c_{i1},...,c_{ij}\}$ and $\mathbf{t}_i^* = \{t_{i1}^*,...,t_{ij}^*\}$. For each value of $\sigma_{RE}$, recorded times were defined as $t_{ij}^* = t_{ij} + \delta$, where $\delta \sim \textrm{N}(0, \sigma_{RE})$. This step was repeated $B = 1000$ times per patient, providing a distribution of possible error-prone \mic{} estimates for each value of $\sigma_{RE}$.
\end{enumerate}

After obtaining these estimates, we examined the within patient bias, calculated as
\begin{align} \label{biasEq}
\textrm{bias}_i = \frac{1}{B} \sum\limits_{b=1}^{B} \bigg\{h(\hat{\boldsymbol{\theta}}_i^*)_b - h(\hat{\boldsymbol{\theta}}_i)\bigg\}.
\end{align}

For the second approach, we took a non-stochastic look into the impact of TREs on blood draw and infusion times. Each infusion schedule had set times at which we considered a single blood draw to have been taken, detailed below. Simulations assumed that a blood draw was taken at each point in isolation (i.e., each time point was examined separately).

For 1000 simulated patients and for each of the sampling times, we used the true concentration value from the two-compartment model solution as their data point. However, unlike the first approach, we did not include the model error when generating the observed concentration. This decision isolated the impact of TREs on \mic{} from the general error in the concentration value. For each infusion schedule, we set specific times at which blood draws occurred during the fourth infusion cycle. Those sample times were associated with:

\begin{enumerate}
  \item start of infusion ($t$ = 24h for both infusion schedules),
  \item midpoint of infusion ($t$ = 24.25h or 26h for the short or long infusion schedules, respectively),
  \item end of infusion ($t$ = 24.5h or 28h for the short or long infusion schedules, respectively),
  \item 30 minutes after infusion ($t$ = 25h or 28.5h for the short or long infusion schedules, respectively),
  \item one hour after infusion ($t$ = 25.5h or 29h for the short or long infusion schedules, respectively),
  \item two hours after infusion ($t$ = 26.5h or 30h for the short or long infusion schedules, respectively), and
  \item 15 minutes before the next infusion ($t$ = 31.75h for both infusion schedules).
\end{enumerate}

For each simulated patient and each time listed above, we computed the patient's true concentration at each time based on their PK parameter vector and the two-compartment model solution.

A variety of TREs of fixed magnitude $\delta_f$ were considered when generating $t_{ij}^*$. The values (in minutes) were $\delta_f \in \{\pm 5, \pm 15, \pm 30, \pm 60, \pm 90\}$.  These errors could occur in either the blood draw time or the infusion time. For each combination of simulated individual, infusion schedule, and blood draw time, we estimated \mic{} from the following models:

\begin{enumerate}
  \item \textit{No time error}: Computed using the error-free patient data $\{c_{ij},t_{ij}\}$ and the correct infusion schedule.

  \item \textit{Draw time error}: Computed using the error-prone patient data $\{c_{ij},t_{ij}^*\}$ and the correct infusion schedule.

  \item \textit{Infusion time error}: Computed using the error-free patient data $\{c_{ij},t_{ij}\}$ and the infusion schedule with error-prone times in the fourth dosing period. Both the start and end of infusion were shifted by the same amount $\delta_f$ to keep the infusion duration constant.
\end{enumerate}

Using Equation \ref{biasEq} we assessed the bias in model estimates by computing the differences between the error-prone model estimates from steps 2 and 3 above, and the error-free model estimates from step 1. We present overall bias induced by the TREs, in addition to examining differential effects of positive/negative error and short/long infusion schedules.

\subsection{Objective 1 results}

A total of 1000 simulated individuals were used in investigating sensitivity of model estimates to TREs. First, each patient had one simulated blood draw for each of the five dosing windows, sampled uniformly within each dose window. Patient-specific mean absolute bias was computed based on multiple samples from a recording error distribution for each $\sigma_{RE}$. Results within this section are presented across all patients, i.e., summaries of patient-specific means. Across all number of blood draws and all recording error distribution magnitudes, the mean [SD] bias in \mic{} estimates was slightly higher for the 30-minute infusion when compared to the 4-hour infusion (0.105 [0.082] versus 0.084 [0.078], respectively). On a relative scale, this translates to a substantial percent change in bias, particularly when the magnitude of the TRE is large (Table \ref{tab:recerr_pct_re}). For a single blood draw, the 4-hour infusion had a higher mean absolute bias in for draws obtained not during an active infusion (0.09 compared to 0.079) but a slightly lower mean absolute bias for draws during an infusion (0.139 versus 0.165). When there were two or more blood draws, the 4-hour infusion consistently had lower mean absolute bias than the 30-minute infusion, regardless of whether or not the draws were during an active infusion.

\begin{table}
\onehalfspacing
\caption[Median (IQR) percent change in absolute bias of \mic{} estimates as a function of recarding error magnitude.]{Median (IQR) percent change in absolute bias of \mic{} estimates as a function of recarding error magnitude. \label{tab:recerr_pct_re}}
\begin{tabular}{l|cccc} 
&  \multicolumn{4}{c}{Value of $\sigma_{RE}$}  \\
Infusion Schedule &  0.25 & 0.5 & 1.0 & 1.5 \\
\hline
 30 minute   & 10.1 (5.4, 21.3) & 20.9 (11.6, 40.8) & 39.1 (23.2, 65.3) & 52.0 (32.8, 79.0)  \\
 4 hour    & 7.86 (4.19, 16.8) & 15.0 (8.13, 32.6) & 26.8 (14.8, 57.3) & 36.0 (19.5, 70.2) \\
\hline
\multicolumn{5}{L{0.9\textwidth}}{
  Results are presented as Median [IQR] for the within-individual percent absolute bias in \mic{} for varying values of the recording error standard deviation $\sigma_{RE}$. Percent absolute bias is computed relative to the individual's model-estimated value of \mic{} when no TRE is present. Simulated patients with an estimate equal to 0 are excluded from this summary as percent change is undefined in this instance. Note that this means the summaries for the 4 hour infusion in particular may be slight under-estimates, as a larger proportion of simulated individuals have an \mic{} estimate of 0.
  }
\end{tabular}
\end{table}

\begin{figure}
    \centering
    \caption[Bias in estimates of \mic{} with recording errors in blood draw time by recording error magnitude.]{Bias in estimates of \mic{} with recording errors in blood draw time by recording error magnitude.}
    \includegraphics[width=5.5in]{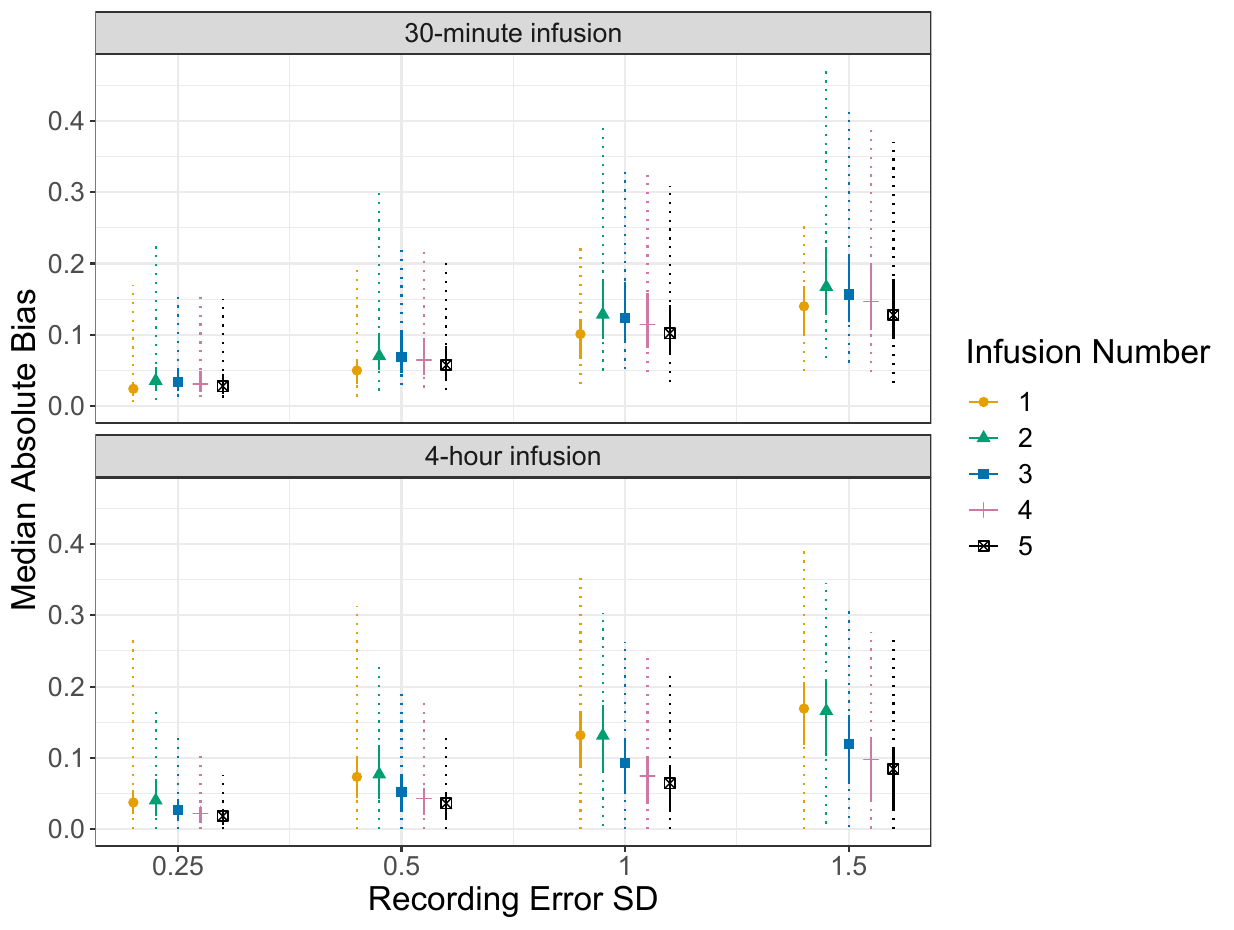}

    \begin{minipage}[H]{\textwidth}
    \singlespacing
    \footnotesize
    Bias in \mic{} between model estimates computed with and without time recording error for $\sigma_{RE}$ = 0.25, 0.5, 1.0, and 1.5. Results for infusion number $k$ are based on the cumulative information from $k$ blood draws up through the end of that dosing window. Points and solid lines represent the median [IQR] of patient-specific summaries across 1,000 simulated individuals; dotted lines represent intervals bounded by the 2.5$^{th}$ to 97.5$^{th}$ percentiles.
    \end{minipage}
    \label{fig:recerrREsd}
\end{figure}

As the SD of the recording error distribution increased, the magnitude of overall bias increased (Figure \ref{fig:recerrREsd}). Outside of this, there were no notable differences observed across the number of blood draws as a function of $\sigma_{RE}$. For all recording error distribution sizes, the bias in \mic{} estimates for the 4-hour infusion started higher for a single blood draw and steadily decreased with each additional sample obtained. For the 30-minute infusion, the absolute bias initially increased from the first to the second blood draw, before reducing with each additional blood draw.

\begin{sidewaysfigure}
\centering
\caption[Bias in model estimates with and without recording errors in blood draw time.]{Bias in model estimates with and without recording errors in blood draw time}
\includegraphics[width=7.5in]{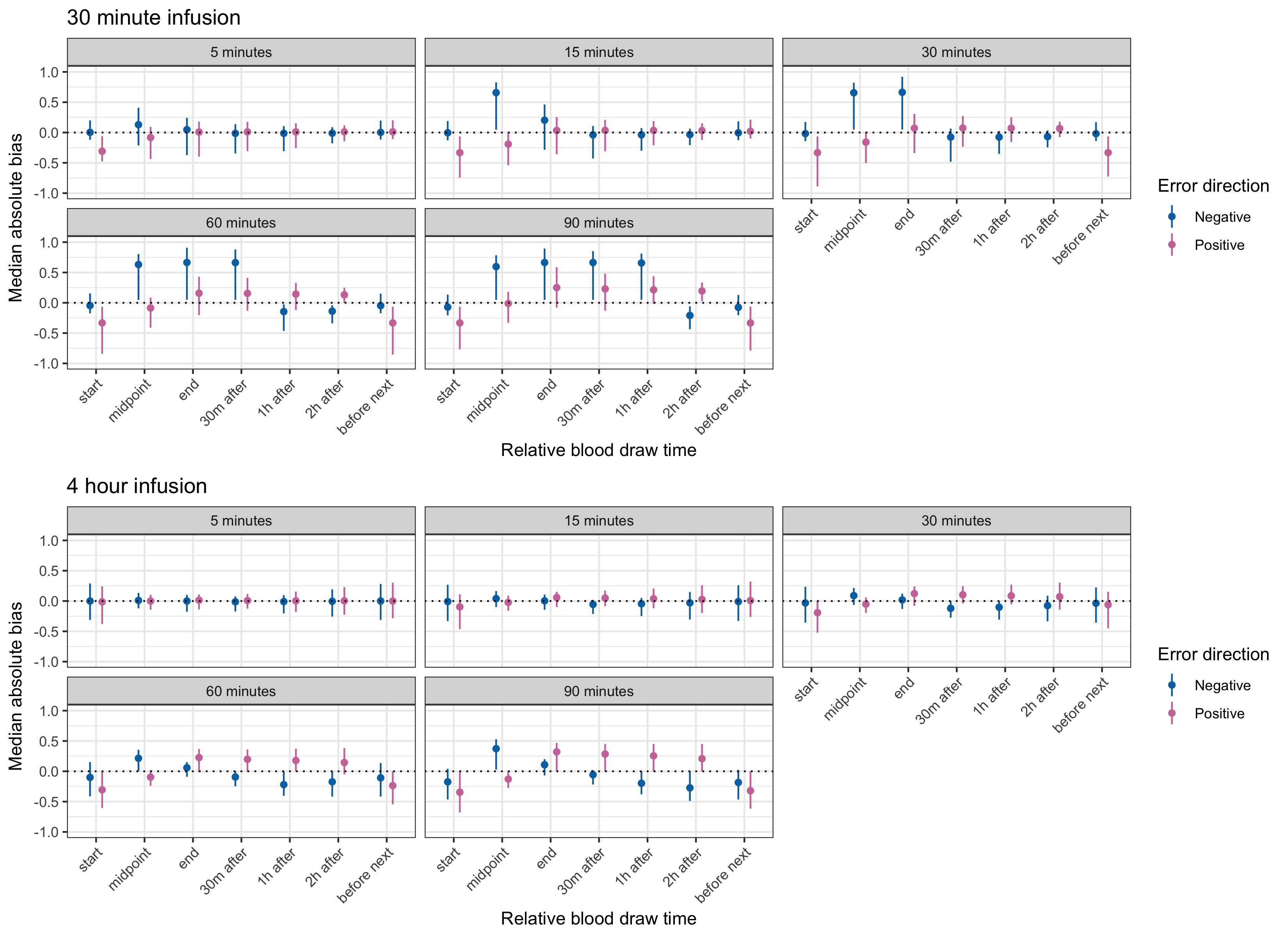}

 \vspace{-0.5\baselineskip}
\begin{minipage}[H]{\textwidth}
    \singlespacing
    \footnotesize
    Bias in \mic{} estimates between model with and without time recording error. Each individual had a patient-specific bias in \mic{} for various error values $\delta_f \in \{\pm 5, \pm 15, \pm 30, \pm 60, \pm 90\}$. Points represent the median of these patient-specific summaries across 1000 simulated individuals; solid lines represent intervals bounded by the 2.5$^{th}$ to 97.5$^{th}$ percentiles.
    \end{minipage}

\label{fig:posneg}
\end{sidewaysfigure}

Next, recording error magnitudes $\delta_f$ were fixed rather than randomly generated and blood draw times occurred at specific times relative to infusion start time. This simulation additionally summarized bias in \mic{} as a result of errors in the infusion time.

Without considering the magnitude or direction of the recording error the 4-hour infusion schedule had lower mean absolute bias (0.131 and 0.117 for blood draw and infusion time errors, respectively) than the short (30 minute) infusion schedule (0.235 and 0.214 for blood draw and infusion time errors, respectively). Stratifying by the sign of the recording error, we see that for the long infusion schedule, draw time errors in opposite directions appear to have opposing effects on the bias of estimates of \mic: negative errors lead to positive bias and positive errors lead to negative bias (Figure \ref{fig:posneg}). With the short infusion schedule, these patterns were less apparent. Positive and negative recording errors could induce bias in the same direction. Similar patterns are seen when recording errors occur in the infusion time.

\begin{figure}[!hb]
\centering
\caption[Distribution of absolute bias in model estimates for errors in blood draw times and infusion times.]{Distribution of absolute bias in model estimates for errors in blood draw times and infusion times. \label{fig:ptile}}
\includegraphics[width=5in]{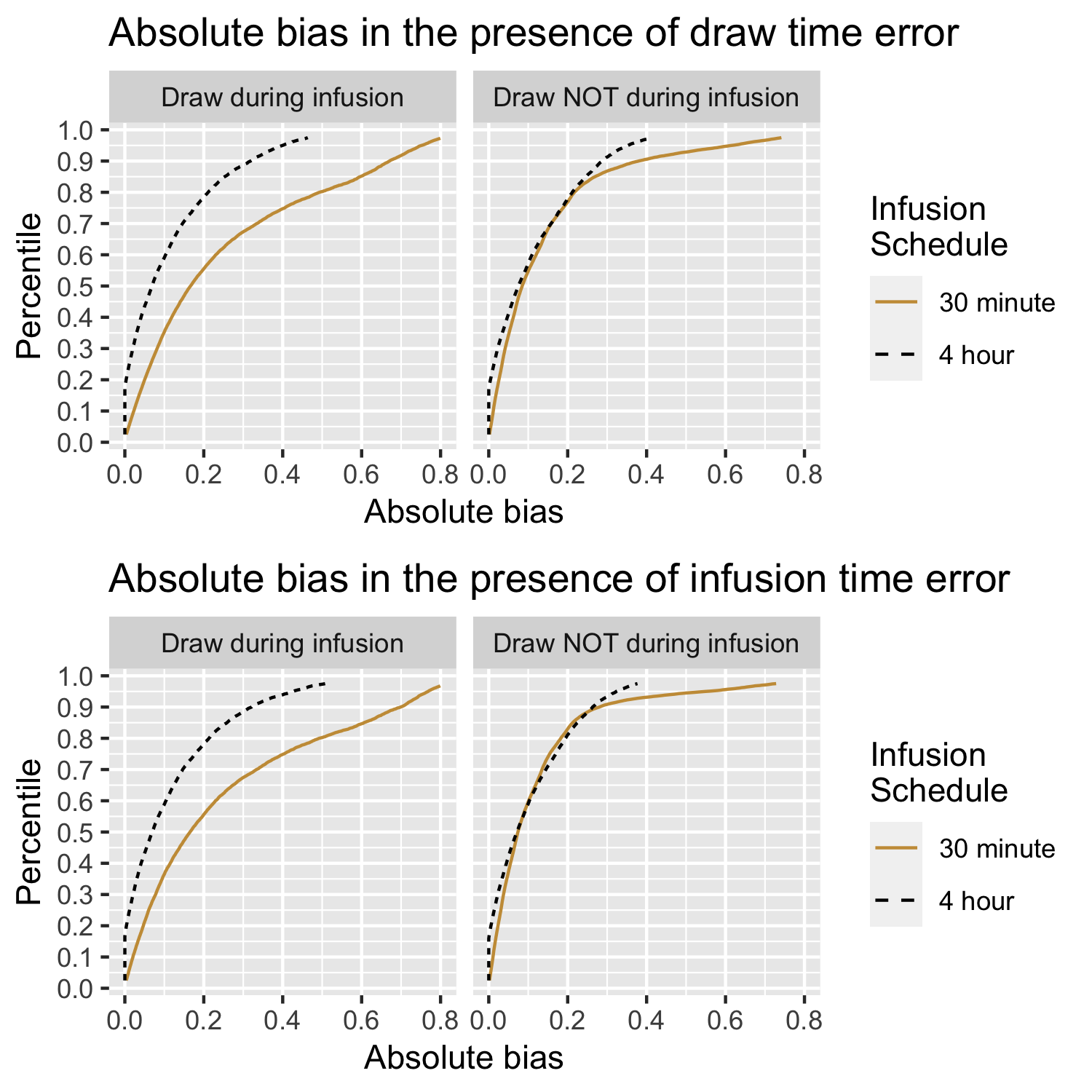}

\begin{minipage}[H]{\textwidth}
    \singlespacing
    \footnotesize
    Empirical cumulative distribution functions for the percentiles of absolute bias in model estimates of \mic{}. Results are presented for both errors in the blood draw time (top panel) and errors in the infusion start/end times (bottom panel). Bias is stratified and considered separately depending on whether or not the blood draw was taken during an active infusion period.
    \end{minipage}
\end{figure}

We previously mentioned that the long infusion schedule appeared to be more robust to both blood draw and infusion time errors overall. This is more comprehensively laid out when viewing empirical cumulative distribution functions for model bias (Figure \ref{fig:ptile}). The empirical distribution of biases induced by both blood draw and infusion time errors were similar between the 30-minute and 4-hour infusion schedules. Whether or not the corresponding blood draw used to estimate \mic{} was taken during an active infusion did appear to differentially impact bias based on infusion duration. If the blood draw was taken during infusion, then bias was much higher (around 0.2 on average) for the short infusion schedule than for the long infusion schedule. This finding further supports that the long infusion schedule is more robust to these types of errors.

\section{Objective 2: Policy change simulations}

\subsection{Objective 2 methods}

Driven by the demonstrated impact of TREs from Objective 1, we examined policy changes to potentially mitigate the impact of these errors. Specifically, we were interested in changes in sampling design that are minimally disruptive to a clinical workflow, while still providing a reduction in bias of the \mic{} estimate in the presence of recording errors. For the remainder of our simulation studies, we focus solely on recording errors in the blood draw time, as the timing of these tends to be more flexible than the timing of administered doses.

\subsubsection{Strategy (i): Delay to end of infusion}

The first strategy we evaluated was based on our observation that blood draws taken during an active infusion period were associated with higher bias. In this policy, a blood draw taken during infusion was instead delayed to the end of infusion, or to the end of infusion plus some fixed delay $d = 15,\ 30, \text{ or } 60$ minutes. We compared the impact of TREs using either the original mid-infusion blood draw time or the delayed time, measured by the bias in \mic.

We simulated 1000 patients, each with one blood draw per dosing window (i.e., the time from the start of one infusion until just before the start of the next infusion). Each simulated patient $i$ had blood draw times $t_{ij}$ sampled uniformly within each of the five dosing windows: $t_{i,j} \sim \textrm{Unif}(s_j, e^{*}_j)$ for $j=1,...,5$, with $e^{*}_j$ representing the end of the active infusion period. Each time had a corresponding simulated concentration measurement $c_{ij}$ obtained from the two-compartment model solution. For each patient and blood draw we computed the following quantities:

\begin{enumerate}

  \item $h(\hat{\boldsymbol{\theta}}_i)$ = \mic{} using the original error-free draw times $\{c_{ij},t_{ij}\}$, which occurred during active infusion periods.
  \item $h(\hat{\boldsymbol{\theta}}_i^*)$ using the original error-prone draw times $\{c_{ij},t_{ij}^*\}$. For each value of $\sigma_{RE}$, we simulated $B=1000$ values of $t_{ij}^*$, used to generate a distribution of \mic{} estimates based on error-prone times.

  \item $h(\hat{\boldsymbol{\theta}}_{i,delay})$ using delayed error-free draw times. A concentration measurement was simulated in accordance with each delayed blood draw time, and the model estimate was computed based on the error-free times $\{c_{ij,delay};t_{ij,delay}\}$.

  \item $h(\hat{\boldsymbol{\theta}}_{i,delay}^*)$ using delayed error-prone draw times $\{c_{ij,delay};t_{ij,delay}^*\}$. For each delayed blood draw time in step 3, and for each value of $\sigma_{RE}$, we simulated $B=1000$ values of $t_{ij,delay}^*$, used to generate a distribution of \mic{} estimates based on error-prone times.
\end{enumerate}

Based on these estimates, we computed bias as in Equation \ref{biasEq}. We determined whether the bias induced by TREs when the blood draws occurred at or shortly after the end of an infusion (Equation \ref{biasEq} using $h(\hat{\boldsymbol{\theta}}_{i,delay}^*)$ to $h(\hat{\boldsymbol{\theta}}_{i,delay})$) was lower than the bias when blood draws occured during an active infusion (Equation \ref{biasEq} using $h(\hat{\boldsymbol{\theta}}_i^*)$ to $h(\hat{\boldsymbol{\theta}}_i)$).

\subsubsection{Strategy (ii): Best sample time}

With the second strategy, we attempted to mitigate the impact of TREs by empirically identifying times which are more robust to these types of errors. We simulated 1000 patients who received three blood draws, each of which is prone to recording error. In the baseline approach of uniform random timing of samples, blood draw times were sampled as follows:

\begin{enumerate}
  \item First draw during the second or third dose window: $t_{i,1} \sim \textrm{Unif}(8,24)$.

 \item Second draw during the fourth dose window: $t_{i,2} \sim \textrm{Unif}(24, 32)$.

 \item Third draw during the fifth dosing window: $t_{i,3} \sim \textrm{Unif}(32, 40)$.
\end{enumerate}

The best sample time approach also started with the first blood draw being sampled uniformly during the second or third dose window. After this initial sample, we fit our model to obtain the patient's posterior PK parameter estimate $\hat{\boldsymbol{\theta}}_i$. To select the timing of the second blood draw, we considered candidate times in increments of 30 minutes from the start to the end of the fourth dosing period (i.e., $candidate_k \in \{24, 24.5,..., 31.5, 32\}$). The following approach was evaluated for $\sigma_{RE} = \{0.25, 0.5, 1.0, 1.5\}$.

Using the posterior-estimated PK parameter after the first sample, we simulated a concentration measurement at each candidate time. Then, we simulated $B=5$ values of $t_{ij}^* = candidate_k + \delta$, $\delta \sim \textrm{N}(0,\sigma_{RE})$. The relatively low value of B here was due to a trade off for the time required to fit the model multiple times at each candidate time. While a single computation to obtain the \mic{} estimate is quick, as $B$ increases it quickly becomes time-intensive. The next blood draw is assigned to occur at the time where the average bias in \mic{} (again using Equation \ref{biasEq}) between the estimates using recorded times versus the candidate time is lowest. If multiple candidate times yielded the same lowest average bias, then the lowest average bias in predicted concentration was used as a tiebreaker.

This process was then repeated for the third blood draw. The posterior estimate of the patient's PK parameter was updated using data from the first two blood draws. Candidate times 32.5, 33,..., 39.5, 40 were considered within the fifth dose window, and the time which minimized bias in the presence of recording errors was selected as the third and final draw time. To evaluate the mitigation approach, we computed \mic{} for both the baseline and best sample time implementations using on all three blood draws on each patient and computed bias using Equation \ref{biasEq}.

\subsubsection{Strategy (iii): Informed allocation}

With the third strategy, we viewed the problem through a different lens: we assumed that resources were limited and that we could only collect additional samples for a subset of patients. The general structure of this simulation was that all patients receive a single initial blood draw, occurring at a time uniformly sampled from the second or third dosing windows. We assumed that we could obtain a second sample from half of the patients, and used information from their initial samples to make that determination. For this simulation, we considered cohorts of 100 patients (50 with one blood draw, and 50 with two) and repeated the simulation 100 times. We assessed the bias mitigation from TREs within each cohort with or without informed allocation.

For each patient's initial blood draw, we simulated a concentration measurement based on the underlying true sample time. We then fit the model using this concentration measurement, but the error-prone blood draw time to obtain each patient's observed \mic{} estimate and corresponding approximate 95\% credible interval. Since time and resources are assumed to be limited, we determined which patients were in the highest need of obtaining a more accurate \mic{} estimate based on (i) the sample data we have on them so far and (ii) the potential impact of recording errors on samples collected so far.

We consider three different criteria for determining which patients require an additional sample:

\begin{enumerate}
  \item \textit{Random (baseline)}: In this case, the patients who receive a second blood draw are randomly selected.

  \item \textit{Credible interval width}: Wider credible intervals indicate less certainty about the patients' current \mic{} estimates. The patients who received a second blood draw were those about whom we had the most uncertainty after their initial measurement. Second samples were assigned to patients in descending order, starting with the patient who had the widest interval.

  \item \textit{Quadrature-based uncertainty calibration}: The patients receiving a second sample were determined based on their empirical sensitivity to recording errors. Using Gauss-Hermite quadrature with 12 nodes, we simulated TREs and applied each to all patients' initial blood draw times. Each patient's \mic{} estimate was recalculated using sample times based on the 12 recording error magnitudes, and we calculated the quadrature-weighted variance of these estimates. A larger variance indicated that the patient's estimate was more sensitive to TREs at their initial sample time. Patients with the highest \mic{} variance in the presence of TREs were assigned to receive a second sample.

\end{enumerate}

In all cases, if multiple patients had the same criteria (e.g., the same credible interval width for approach 2) and there were not enough samples left to assign to all of them, we randomly selected patients from this ``eligible'' group. For patients selected to receive a second blood draw under each approach, we randomly sampled a time in the fourth or fifth dosing periods.

For each approach, once the subset of patients receiving a second dose was assigned, concentration measurements were simulated at the second blood draw time. Based on all available data for each patient (i.e, one blood draw for half of the patients and two draws for the other half), we computed \mic{} (i) with no TRE in any blood draw times and (ii) with TREs $\delta \sim \textrm{N}(0, \sigma_{RE})$ applied to all times. We evaluated whether either of the informed allocation approaches (2 and 3 above) reduced the average within-individual impact of TREs relative to the baseline approach of random assignment (1 above).

\subsection{Objective 2 results} 

\subsubsection{Strategy (i): Delay to end of infusion}

The first policy change we examined was assuming a blood draw was originally going to take place during an active infusion. Instead we proposed to delay the draw until the end of infusion, or until 15, 30, or 60 minutes post-infusion. For both short and long infusion schedules, there was no notable difference in the impact on bias of estimates based on the dose window during which the blood draw occurred (first dose, second dose, etc.).

\begin{figure}[ht]
    \centering
    \caption[Bias in \mic{} due to time recording error: delay to post-infusion]{Bias in \mic{} due to time recording error: delay to post-infusion}
    \includegraphics[width=5.5in]{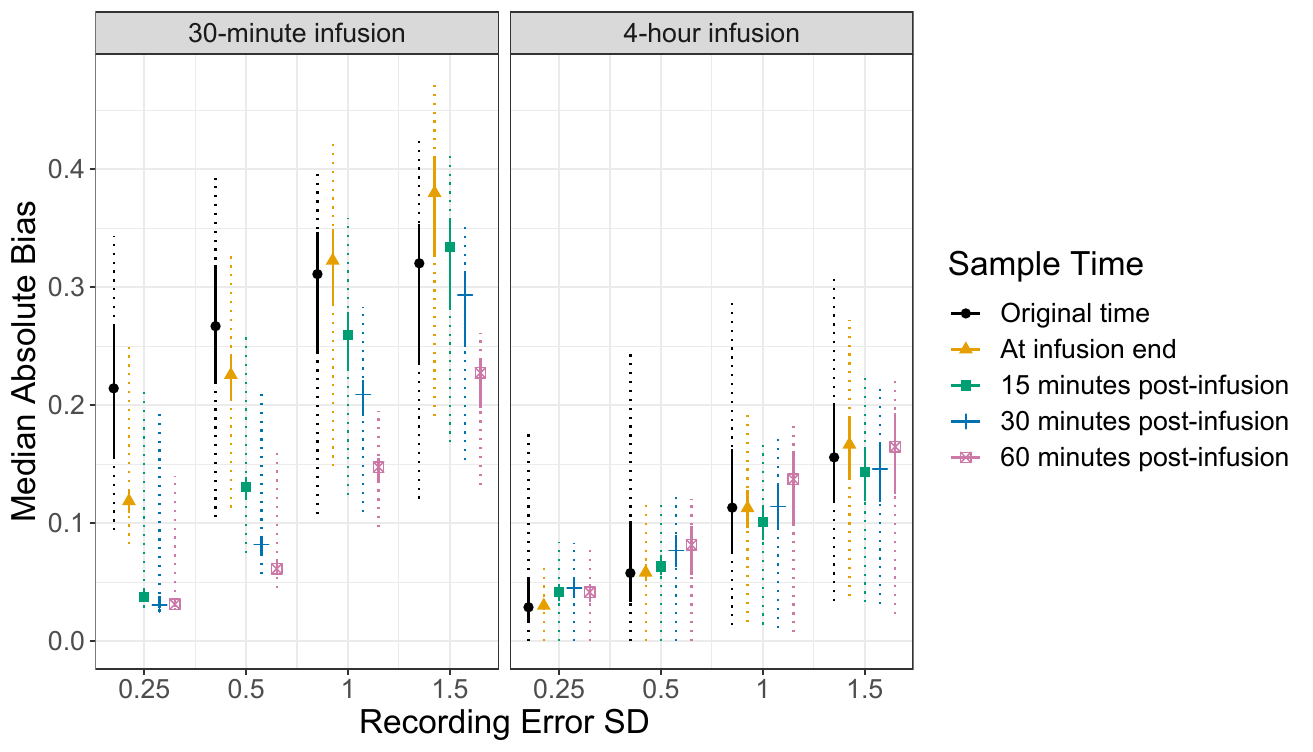}

    \begin{minipage}[H]{\textwidth}
    \singlespacing
    \footnotesize
    Bias in \mic{} estimates between model with and without recording error at original sample time (during an active infusion) or shifted to post-infusion. Each individual had a patient-specific mean bias in \mic{} across the recording error distribution, for $\sigma_{RE}$ = 0.25, 0.5, 1.0, and 1.5. Points and solid lines represent the median [IQR] of these patient-specific summaries across 1,000 simulated individuals; dotted lines represent intervals bounded by the 2.5$^{th}$ to 97.5$^{th}$ percentiles.
    \end{minipage}
    \label{fig:drawshift}
\end{figure}

\begin{table}[!h]
\small
\caption[Median (IQR) percent change in absolute bias of \mic{} estimates at recording error extremes: delay to post-infusion.]{Median (IQR) percent change in absolute bias of \mic{} estimates at recording error extremes: delay to post-infusion. \label{tab:recerr_pct_re_delay}}
\centering
\begin{tabular}{c}
\toprule

\renewcommand{\arraystretch}{1.25}
\begin{tabular}{lcc}
 \multicolumn{3}{l}{30 minute infusion}\\
 & $\sigma_{RE}$ = 0.25 & $\sigma_{RE}$ = 1.5 \\
 \cline{2-3}
 Original time & 64.7 (44.2, 94.9) & 86.6 (59.2, 127.0)  \\
 At infusion end & 36.3 (28.2, 53.4) & 109.0 (73.7, 164.0) \\
 15 minutes post-infusion & 12.6 (10.4, 18.5) & 96.1 (83.9, 18.5) \\
 30 minutes post-infusion & 10.5 (8.9, 15.2) & 83.9 (57.1, 134.0) \\
 60 minutes post-infusion & 10.6 (8.8, 14.9) & 67.0 (47.4, 106) \\
\bottomrule
\end{tabular} \\

\renewcommand{\arraystretch}{1.2}
\begin{tabular}{lcc}
 \multicolumn{3}{l}{4 hour infusion}\\
 & $\sigma_{RE}$ = 0.25 & $\sigma_{RE}$ = 1.5 \\
  \cline{2-3}
 Original time & 8.5 (4.1, 20.2) & 40.1 (23.7, 73.5) \\
 At infusion end & 6.6 (5.1, 13.1) & 37.3 (24.3, 65.6) \\
 15 minutes post-infusion & 9.2 (6.7, 18.5) & 32.1 (20.7, 55.4) \\
 30 minutes post-infusion & 10.1 (7.1, 20.6) & 32.8 (21.1, 59.2) \\
 60 minutes post-infusion & 9.4 (6.5, 19.0) & 40.1 (24.6, 73.8) \\
\end{tabular} \\

\bottomrule

\multicolumn{1}{L{.8\textwidth}}{\footnotesize{Results are presented as Median (IQR) for the within-individual percent absolute bias in \mic{} for the smallest and largest values of $\sigma_{RE}$. Percent absolute bias is computed relative to the individual's model-estimated value of \mic{} when no TRE is present. Simulated patients with an estimate equal to 0 are excluded from this summary as percent change is undefined in this instance. Note that this means the summaries for the 4 hour infusion in particular may be slight under-estimates, as a larger proportion of simulated individuals have an \mic{} estimate of 0.}}
\end{tabular}
\end{table}

For the 30-minute infusion schedule, there was a significant reduction in the bias of the \mic{} estimate when shifting to the post-infusion times (Figure \ref{fig:drawshift}). When the recording error magnitude was assumed to be small, even shifting to the exact end of infusion lowered bias relative to taking the draw during infusion. For larger recording error magnitudes, shifting to the end of infusion resulted in an overall increase in bias of the estimate. Shifting to post-infusion times still resulted in a reduction in bias in these instances, though there was a smaller relative improvement in estimation compared to if the sample had been drawn at the scheduled time.

Unlike the shorter infusion schedule, the 4-hour infusion schedule saw little to no reduction in bias when shifting the time of a blood draw to the end or shortly after the end of infusion (Figure \ref{fig:drawshift}). In fact, large shifts to one hour post-infusion tended to result in a slight increase in bias of the \mic{} estimate. Similar patterns were observed for this infusion schedule for both smaller and larger recording error magnitudes, with the main difference being bias increased as the recording error magnitude increased.

Similar patterns as those observed in Figure \ref{fig:drawshift} are also present when looking at percent change in absolute bias. For small TREs, the 30 minute infusion has a notably lower bias in \mic{} when the blood draw time is delayed until after infusion (Table \ref{tab:recerr_pct_re_delay}). For large TREs, longer delays are necessary to see a decrease, though percent bias remains high even when delaying to one hour after infusion. For the 4 hour infusion, percent bias is relatively stable regardless of how much the blood draw time is delayed.

\subsubsection{Strategy (ii): Best sample time}

The best sample time approach yielded an overall reduction in bias of the \mic{} estimate. Relative to random assignment of subsequent blood draw times, the best sample time method reduced mean (median) absolute bias by 24\% (17\%) for the 30-minute infusion schedule and 9\% (29\%) for the 4-hour infusion schedule (Table \ref{tab:besttime}).

\begin{table}[h]
\centering
\small
\caption[Overall impact of best sample time approach on bias in \mic{} due to time recording error]{Overall impact of best sample time approach on bias in \mic{} due to time recording error \label{tab:besttime}}
\begin{tabular}{llll} \hline
Infusion Duration & Approach          & Mean (SD) Absolute Bias  & Median [IQR] Percent \\
& & & Change in Absolute Bias \\  \hline
 30 minutes & Random time assignment  & 0.069 (0.055) & 18.5 [13.4, 27.2] \\
  & Best sample time       & 0.053 (0.043) & 8.0 [5.8, 12.0] \\
 4 hours & Random time assignment     & 0.053 (0.044) & 7.6 [5.5, 13.7] \\
 & Best sample time           & 0.049 (0.051) & 12.5 [8.8, 24.0]\\
\hline
\end{tabular}
\end{table}

\begin{figure}[!hb]
\centering
    \caption[Impact of recording error distribution and number of blood draws with the best sample time strategy]{Impact of recording error distribution and number of blood draws with the best sample time strategy}
{\includegraphics[scale=0.5]{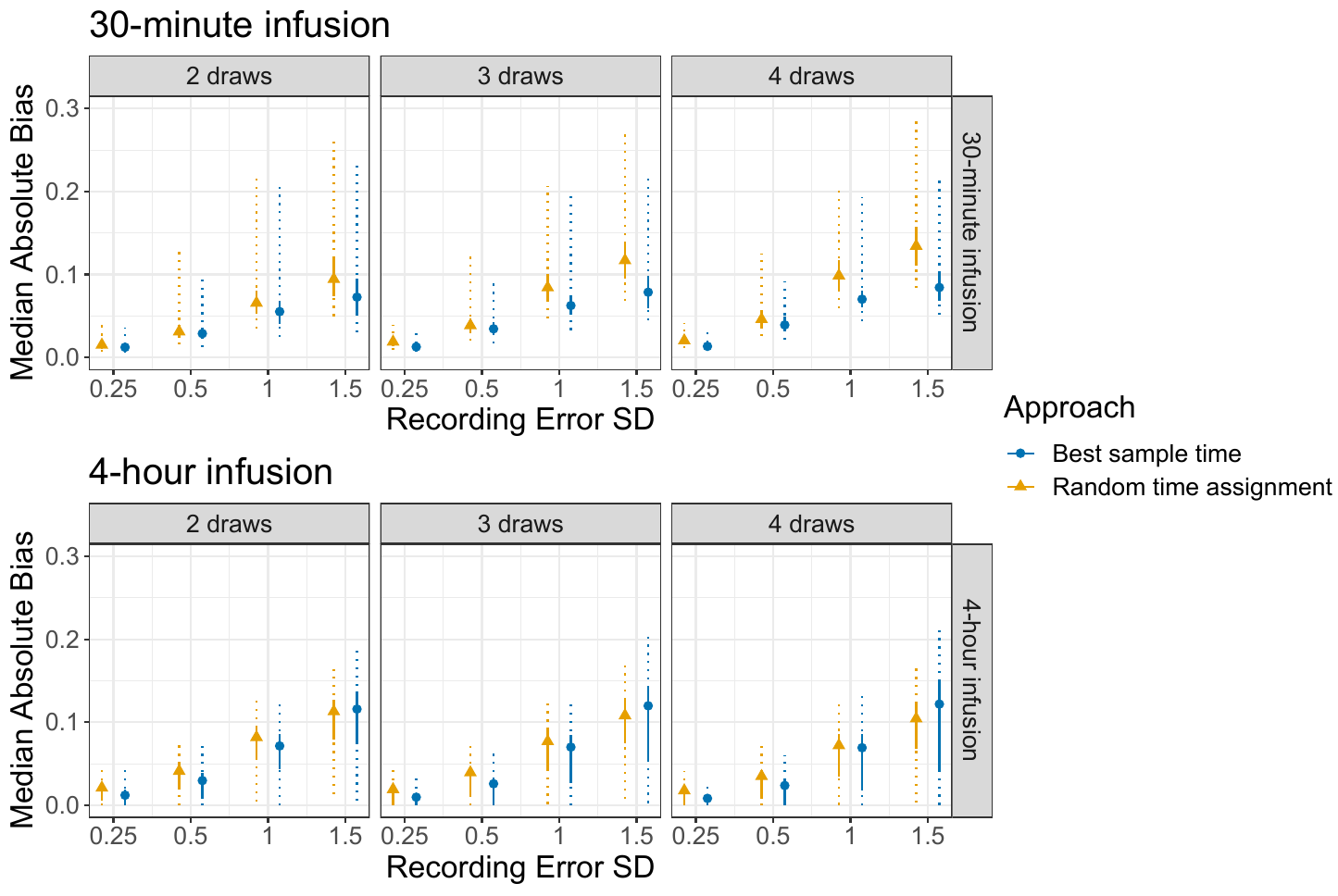}}

    \begin{minipage}[H]{\textwidth}
    \singlespacing
    \footnotesize
    Summary of absolute bias in \mic{} when comparing the baseline approach (random assignment of next blood draw time) to the best sample time approach. Results are stratified by infusion schedule (either 30 minutes or 4 hours) and by the SD of the recording error distribution. Simulated patients received 4 blood draws in total, where draws 2 through 4 were selected differently based on approach. Results for each draw number are based on the bias in \mic{} induced by time recording error up to and including that blood draw measurement.
    \end{minipage}
    \label{fig:besttime_resd}
\end{figure}

Figure \ref{fig:besttime_resd} provides a more in-depth look at the impact of both TRE distribution and number of blood samples available. Results for the first blood draw are not shown, since only the second through fourth were selected via the best sample time or random assignment approaches. For the 30-minute infusion schedule, the median absolute bias in \mic{} increased as the number of error-prone sample draws increased, regardless of the time selection approach. The 4-hour infusion schedule did not have this issue. Bias decreased as the number of blood draws increased when recording error magnitudes were small, and remained stable or increased only slightly for larger recording error magnitudes.

\begin{figure}[!h]
\centering
    \caption[Candidate time selected by best sample time strategy: 30-minute infusion]{Candidate time selected by best sample time strategy: 30-minute infusion}
{\includegraphics[scale=0.5]{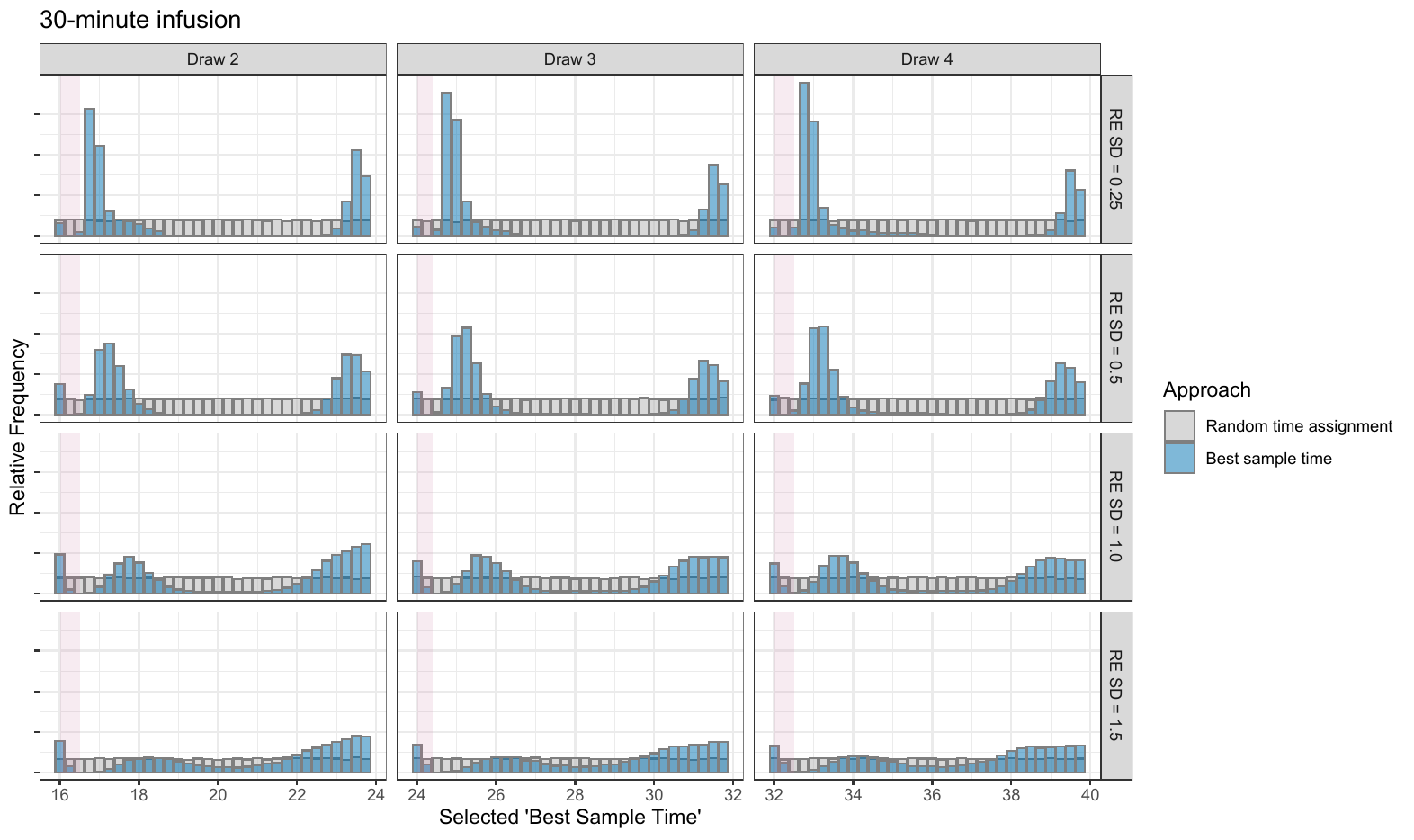}}

    \begin{minipage}[H]{\textwidth}
    \singlespacing
    \footnotesize
    Times selected for 30-minute infusion as the best candidate blood draw time based on minimizing bias in \mic{} statistic due to time recording error. Red shaded region indicates the period of active infusion within each dose window.
    \end{minipage}
    \label{fig:besttime_drawtimes30m}
\end{figure}

\begin{figure}[!ht]
\centering
    \caption[Candidate time selected by best sample time strategy: 4-hour infusion]{Candidate time selected by best sample time strategy: 4-hour infusion}
{\includegraphics[scale=0.5]{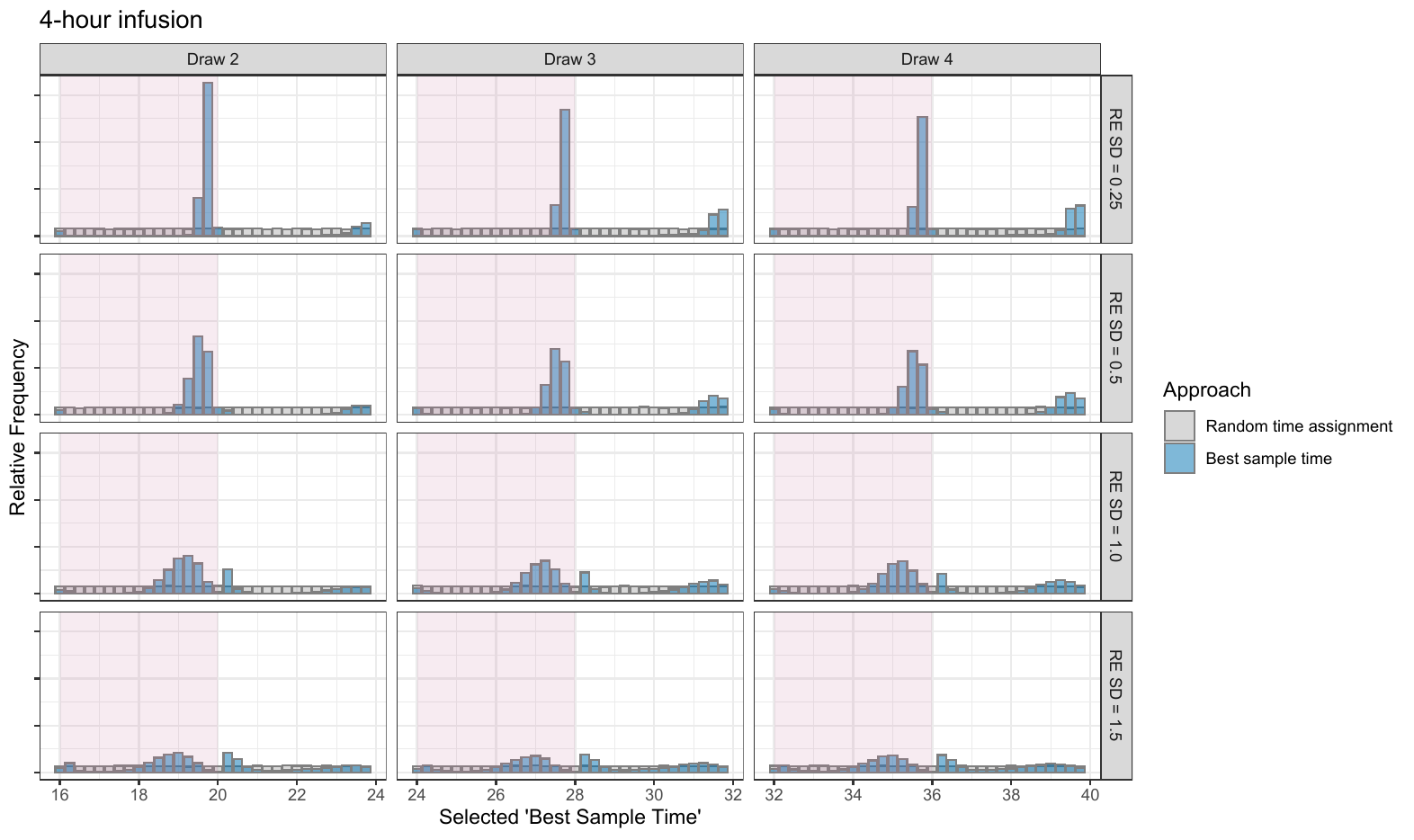}}

    \begin{minipage}[H]{\textwidth}
    \singlespacing
    \footnotesize
    Times selected for 4-hour infusion as the best candidate blood draw time based on minimizing bias in \mic{} statistic due to time recording error.  Red shaded region indicates the period of active infusion within each dose window.
    \end{minipage}
    \label{fig:besttime_drawtimes4h}
\end{figure}

When recording error magnitudes were smaller, the best sample time approach performed similarly to random assignment of next sample time regardless of infusion schedule. For larger errors, however, the 30-minute infusion had lower median absolute bias when using the best sample time approach. This same pattern did not hold true for the 4-hour infusion schedule, where bias in \mic{} under the best sample time approach was similar to random assignment, or in the case of the largest error magnitudes was occasionally worse than random assignment.

Figures \ref{fig:besttime_drawtimes30m} and \ref{fig:besttime_drawtimes4h} summarize the actual times selected by each approach. The 30-minute infusion favored times that were either one or two hours after an infusion ended or shortly before the next infusion was about to begin. As recording error magnitude increased, the selection frequency of times in the middle of the dose window (around three to six hours after the infusion start time) increased. The 4-hour infusion schedule strongly favored selecting times that were shortly before the end of an active infusion window, with a smaller secondary preference for times just before the start of the next infusion. Larger recording error magnitudes led to increases in selected times in the one to two hours before and two to four hours after infusion end.

\subsubsection{Strategy (iii): Informed allocation}

\begin{figure}[!h]
\centering
    \caption[Bias in \mic{} due to time recording error: informed allocation]{Bias in \mic{} due to time recording error: informed allocation}
{\includegraphics[scale=0.8]{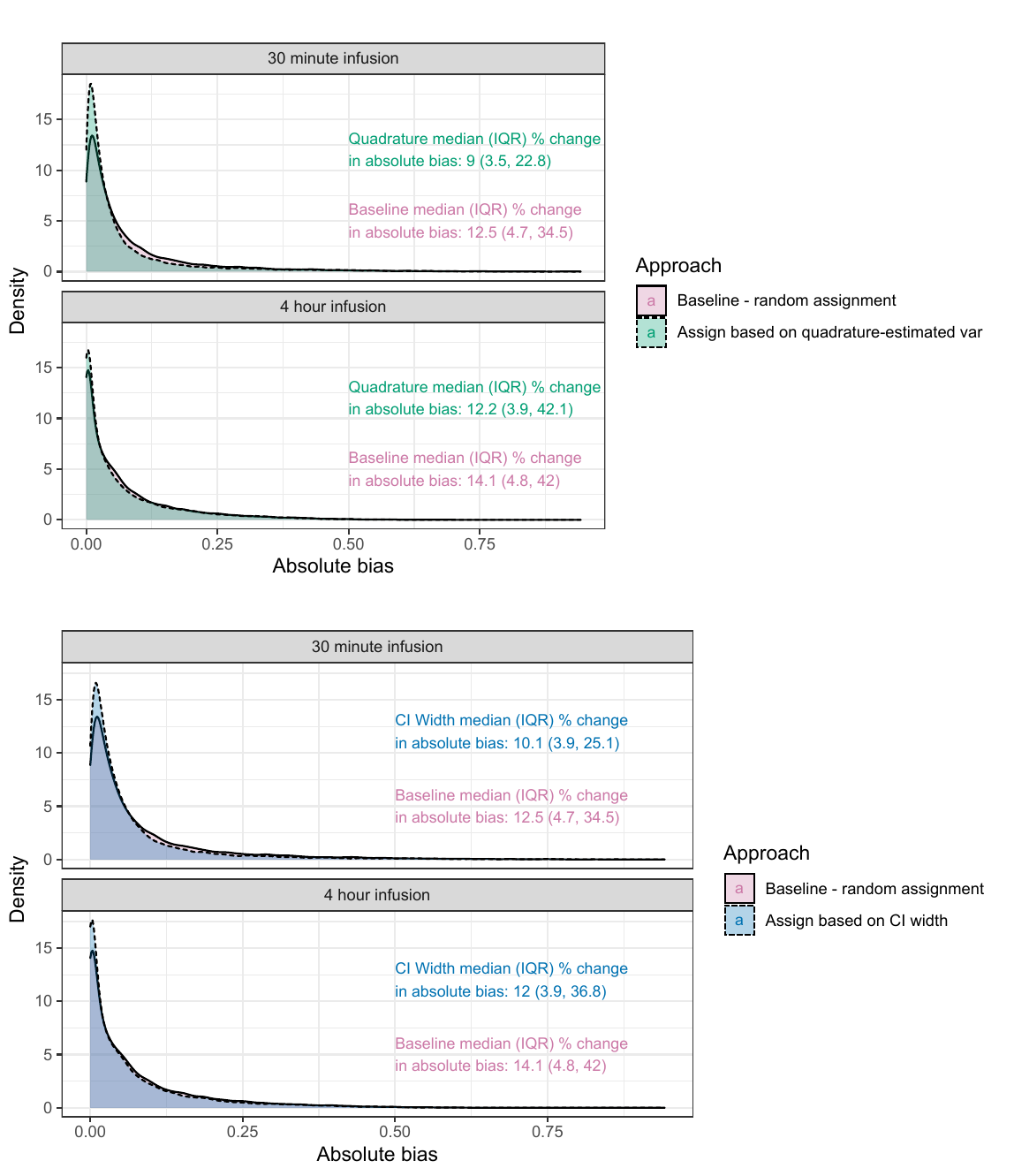}}

    \begin{minipage}[H]{\textwidth}
    \singlespacing
    \footnotesize
    Overlaid distributions of absolute bias in \mic{} between different blood draw allocation strategies, averaged across small to medium error magnitudes $\sigma_{RE}$ = 0.25, 0.5. Median (IQR) percent change in absolute bias relative to model estimates with no TRE are presented as an additional summary.
    \end{minipage}
    \label{fig:infall_density}
\end{figure}

Informed allocation simulations focused on reducing overall bias by using information from a single blood draw to inform which patients were most in need of additional data assuming the presence of TREs. Based on a set list of candidate blood draw times, we compared baseline (random selection) with assigning additional draws based on individuals with the widest credible interval width in their \mic{} estimate (i.e., most uncertainty) or assigning time based on quadrature-estimated ``robustness'' to TREs. Figure \ref{fig:infall_density} describes the absolute bias in \mic{} in the presence of TREs across these scenarios.

Bias was generally low across all scenarios, though we noticed a small reduction when using informed allocation strategies. The difference in median absolute bias between informed strategies reflect a 9$-$15\% reduction in bias for the credible interval-based approach across infusion schedules. For the quadrature-based approach, there was an approximately 10\% reduction in bias for the 30-minute infusion schedule, and a 6$-$7\% reduction for the 4-hour infusion schedule.

\section{Discussion}

Across all simulation studies performed, the 4-hour infusion schedule consistently demonstrated more robustness in the presence of TREs when estimating \mic. This was especially true for estimates based on blood draws taken during an active infusion, where the 30-minute infusion schedule had significantly higher bias. This aligns with expectations based on prior studies that noted areas of high nonlinearity increased bias in estimates of pharmacokinetic parameters \citep{choi2013}. The slower infusion rate of the 4-hour schedule resulted in much slower change in concentration over time when compared with the active infusion period for the 30-minute schedule.

The initial characterization of the impact of TREs showed that for a single blood draw, the 30-minute infusion schedule appeared to have notably lower mean absolute bias in \mic{} than the 4-hour infusion. This may be related to the fact that samples taken during an infusion are more impacted by TREs. For both infusion schedules, draws taken during infusion were associated with higher bias in \mic. For the 4-hour infusion schedule, this problem was both far less impactful (on average resulting in a 15\% increase in bias relative to draws not during an infusion, compared with a 75\% increase for the 30-minute infusion) and far more prevalent when uniformly sampled (about 50\% of draws compared with about 6\% of draws). When estimating \mic{} based on a single blood draw, the 4-hour infusion schedule suffered more from this higher rate of draws taken during infusion. As more blood draws were obtained, however, it may be that the 30-minute infusion had more difficulty ``recovering'' from the cumulative impact of multiple times with recording error, whereas the 4-hour infusion was generally more robust to these types of errors to begin with.

Since the 4-hour infusion schedule was overall less susceptible to these types of errors, mitigation strategies seemed less helpful for this slower, more gradual infusion schedule. One example was our suggested approach of delaying the blood draw time to the end of or shortly after infusion. While this strategy resulted in lower bias in \mic{} estimation when shifting to post-infusion time for the 30-minute infusion schedule, there was little to no benefit observed for 4-hour infusion schedule. In fact, in some settings with the highest recording error magnitudes, delaying the time of blood draw was slightly worse. Since the infusion itself is so gradual and at a fixed rate by design, it iss possible that the elimination rate in the period shortly after an infusion ends may temporarily be greater than the observed infusion rate. Therefore, it may not be worth the inconvenience of returning at a later time to take the sample for longer duration infusion schedules.

Across multiple simulations, the longer infusion schedule either benefited or saw no negative impact from additional sample data. The shorter infusion, however, often had a positive correlation between bias in \mic{} estimation and the number of blood draws available. This provides more evidence that the cumulative impact of multiple error-prone sample times has a stronger impact on short infusion schedules with periods of rapid change in drug concentration. In these situations, it may be more important to implement changes to data collection to minimize bias in estimation of clinical endpoints from time errors.

The best sample time approach was another strategy that had a larger impact on reduction in \mic{} bias for the shorter 30-minute infusion schedule, especially for large recording error magnitudes. This approach also has the disadvantage of being significantly slower to implement than other approaches, taking approximately 15 minutes to estimate \mic{} for five randomly sampled recording errors across multiple candidate sample times. If one wanted to increase the efficacy of this strategy by increasing number of simulated recording error values, that would further increase implementation time. Thus, the decision of whether or not to utilize this strategy should consider the trade off between computational efficiency and error calibration accuracy.

Informed allocation simulations indicated that prioritization strategies for deciding \textit{how} to collect an additional blood draw for rather than \textit{when} can also be helpful in reducing the average impact of TREs across individuals. With no mitigation strategy, longer infusions are generally a better option where possible when TREs are an important concern. In cases where shorter infusion schedules are being used, the strategies presented in this paper can be used to modify the collection of blood samples and minimize the impact of these errors up-front.

The differences in the success of the strategies proposed in this paper between the 30-minute and 4-hour infusion schedules indicate that the rate of drug infusion or absorption is a driving factor in the impact of TREs on bias of estimates. Particularly for drugs that must be administered more rapidly, design mitigation strategies are an additional tool that can be used to improve statistical estimation through higher quality input data. Future work in this area could build upon the results in this paper to evaluate the impact of combining both design mitigation strategies and ME-based statistical models to further reduce the impact of these errors.

\section{Conclusion}

In this chapter we investigated the susceptibility of an approximate Bayesian pharmacokinetic model to TREs in blood draw samples. Our results both corroborate existing findings in statistical and clinical literature, as well as expand on this knowledge by summarizing differential effects for infusion schedules of varying duration. We proposed and evaluated various mitigation strategies in design and data collection to improve the quality of opportunistic sample data to be more robust to TREs. The strategies proposed here are minimally disruptive and could be implemented in a clinical setting to improve data collection for patients during the course of inpatient treatment. These findings also demonstrate that all strategies are not equally effective for all dosing designs, even when dealing with the same drug.

\section{Acknowledgements}
This work was funded in part by CTSA award No. TL1TR000447 from the National Center for Advancing Translational Sciences. Contents of this work are solely the responsibility of the authors and do not necessarily represent the official views of the National Center for Advancing Translational Sciences or the National Institutes of Health.

\bibliographystyle{unsrtnat} 
\bibliography{references}

\end{document}